\documentclass[aps,pre,twocolumn,showpacs]{revtex4-1} 
\usepackage{color,soul}
\usepackage{graphicx}
\usepackage{siunitx}
\usepackage[breaklinks=true,colorlinks=true,linkcolor=blue,urlcolor=blue,citecolor=blue]{hyperref}

\usepackage{amsmath, amsthm, amssymb}
\usepackage{amsthm}
\usepackage{multirow}
\usepackage[normalem]{ulem}
\usepackage{soul}
\usepackage{makecell}
\usepackage[table]{xcolor}
\definecolor{Gray}{gray}{0.85}
\definecolor{LightCyan}{rgb}{0.88, 1, 1}
\definecolor{Apricot}{rgb}{0.98, 0.81, 0.69}
\usepackage{threeparttable}
\newcommand{\be}{\begin{equation}}
\newcommand{\ee}{\end{equation}}
\newcommand{\bea}{\begin{eqnarray}}
\newcommand{\eea}{\end{eqnarray}}

\usepackage[rightcaption,]{sidecap}
\sidecaptionvpos{figure}{c}

\begin{document}

\title{
Two-dimensional squishy glass: yielding under oscillatory shear}
\author{Sayantan Ghosh}
\email{sayantang@imsc.res.in}
\affiliation{The Institute of Mathematical Sciences, C.I.T. Campus,
Taramani, Chennai 600113, India}
\affiliation{Homi Bhabha National Institute, Training School Complex, Anushakti Nagar, Mumbai, 400094, India}
\author{Rahul Nayak}
\email{rahulnayak@imsc.res.in}
\affiliation{The Institute of Mathematical Sciences, C.I.T. Campus,
Taramani, Chennai 600113, India}
\affiliation{Homi Bhabha National Institute, Training School Complex, Anushakti Nagar, Mumbai, 400094, India}
\author{Satyavani Vemparala}
\email{vani@imsc.res.in}
\affiliation{The Institute of Mathematical Sciences, C.I.T. Campus,
Taramani, Chennai 600113, India}
\affiliation{Homi Bhabha National Institute, Training School Complex, Anushakti Nagar, Mumbai, 400094, India}
\author{Pinaki Chaudhuri}
\email{pinakic@imsc.res.in}
\affiliation{The Institute of Mathematical Sciences, C.I.T. Campus,
Taramani, Chennai 600113, India}
\affiliation{Homi Bhabha National Institute, Training School Complex, Anushakti Nagar, Mumbai, 400094, India}
\date{\today}
\begin{abstract}

The yielding response to an imposed oscillatory shear is investigated for a model two-dimensional dense glass composed of bidisperse, deformable polymer rings, with the ring stiffness being the control parameter. In the quiescent glassy state, the more flexible rings exhibit a broader spectrum of shape fluctuations, which becomes increasingly constrained with increasing ring stiffness. Under shear, the highly packed rings yield, i.e. the thermal assembly looses rigidity, with the threshold yield strain increasing significantly with decreasing ring stiffness. Further, the rings display significant deviations in their shape compared to their unsheared counterparts. This study provides insights into the interplay between shape changes and translational rearrangements under shear, thus contributing to the understanding of yielding transition in densely packed, deformable polymer systems.

\end{abstract}

\maketitle

\section{Introduction}\label{SEC-1}

Soft amorphous solids, which include foams, emulsions, granular materials, and colloidal suspensions~\cite{katgert2010jamming,pan2023review,vlassopoulos2014tunable,reichhardt2014aspects,behringer2018physics,cohen2014rheology,ji2020interfacial}, are characterized via the existence of yield thresholds in stress or strain~\cite{coussot2014yield,bonn2017yield}, the advantage of which is utilized in diverse material applications. Many of these soft and disordered materials, often termed as {\em squishy matter}, ranging from biological tissues~\cite{ilina2020cell,angelini2011glass,bi2014energy} to geological formations~\cite{kostynick2022rheology,cunez2024particle}, consist of elements that can undergo significant shape fluctuations. This deformability presents challenges in understanding how microscopic geometrical variations influence the material's transition from a rigid to a fluid-like state under mechanical deformation~\cite{manning2023essay}. The mechanical behavior of these materials under deformation is heavily influenced by the characteristics of their constituent particles, including morphology, size, shape, and the interactions between them~\cite{zhao2023role,behringer2018physics,cunez2024particle,gray2018particle,van2009jamming}.  
In systems with such highly deformable objects, low stiffness facilitates extensive deformations, enabling compression beyond the point where typical rigid materials would yield, leading to higher packing densities.  Studying yielding and rigidity transitions in these systems is challenging due to the complex deformations in both particle shapes and packing structures, which significantly affect inter-particle contacts and consequently alter the mechanical properties of the material. Recent studies have demonstrated fluidization via shape changes which is significant in the context of cellular processes~\cite{bi2015density}, whereafter universalities across have been suggested \cite{sadhukhan2022origin} and extended to granular mimics \cite{arora2024shape}.
Understanding the physics of yielding in these systems is crucial for applications where controlling the mechanical response is essential, such as in biomedical engineering, materials science, and geophysics. Studies on yielding continue to shed light on the factors that influence this complex behavior, providing insights that are vital for designing materials with tailored mechanical properties. In this work, we investigate the shear response of a soft glassy system consisting of such deformable objects.  

Our work is motivated by a recent experimental study~\cite{poincloux2024squish}, where compressible rubber rings were used as an experimental model to investigate the effects of large grain deformations on the rheology of disordered media. Probing the response to an oscillatory shear,  a solid-fluid transition was demonstrated for the {\em squish jammed} system, driven by variations in shear amplitude and density, where significant deformations allowed the assembly to absorb shear stress without inducing structural rearrangements in the solid phase. Previously, experimental studies of yielding of colloidal glasses and other soft glassy materials  under applied oscillatory shear have been extensively reported; e.g. see Ref. ~\cite{bonn2017yield, koumakis2013complex, hima2014experimental}. More recently, there have been a series of numerical studies probing the yielding of model atomistic systems to cyclic or oscillatory shear, which has revealed a fascinating phenomenology with features like history dependence, memory formation, shear-banding etc., the underlying mechanisms of which continue to be investigated in details; e.g., see Ref.~\cite{fiocco2013oscillatory, priezjev2013heterogeneous, regev2013onset, leishangthem2017yielding, mungan2021metastability, yeh2020glass, liu2022fate, kumar2022mapping, parley2022mean}. Our work is a first step in utilizing this analysis to flexible objects where deformability provides an additional degree of freedom.

Modelling of soft deformable materials is a challenge \cite{manning2023essay}. Quite often, ultrasoft potentials are  employed to model interactions in systems where particles experience much weaker repulsion compared to traditional models like Lennard-Jones or hard-sphere systems~\cite{miyazaki2019slow}. These potentials allow significant particle overlap without the rapid energy increase observed in more conventional interactions, making them essential for accurately simulating soft matter systems, such as polymers, where interactions are predominantly entropic rather than repulsive. However, at very high densities, ultrasoft potentials may become less effective, particularly in capturing the flexibility and deformation of particles, and therefore modelling via ring polymers or polygons have been proposed ~\cite{boromand2018jamming, gnan2019microscopic,gnan2021dynamical}. Such modelling represent non-confluent systems, unlike vertex models or the cellular Potts model which mimic foams, confluent tissues etc~\cite{okuzono1995intermittent, chen2007parallel, maree2007cellular, alt2017vertex}. In our work, we consider a dense two-dimensional assembly of ring polymers to model the soft amorphous solid~\cite{ghosh2024onset}. Further, unlike the athermal experimental study that motivates our work, we choose to work under thermal conditions, which approximates for a noisy environment that is typical to many soft colloidal or biological systems. We examine a range of ring stiffness values to understand the relationship between stiffness, deformation, and the mechanical response of the system, particularly focusing on how these factors influence the yielding under shear. 

Note that dimensionality plays a crucial role in the glassy behavior observed in thermal ring polymers, particularly in 2D systems where certain dynamic processes, such as threading and looping that are possible in 3D, are absent~\cite{vsiber2013many,miller2011two,roy2022effect,boromand2018jamming}. These differences fundamentally would alter the mechanical response and glassy dynamics of the system. Studies have shown that 2D glasses exhibit unique mechanical properties under compression, including a pronounced sensitivity to particle shape and interaction potential. Notably, 2D systems can exhibit a novel "re-entrant melting" at very high packing fractions, driven by the deformability of the rings ~\cite{kapfer2015two,zhang2009thermal,gnan2019microscopic,gnan2021dynamical,ghosh2024onset}. 
Investigating the yielding process in such two dimensional systems is critical for applications in materials science, where 2D materials are increasingly being used, and understanding their mechanical limits is essential for their reliable integration into devices and structures.

The manuscript is organized as follows. After the introductory discussion in Section I, the details of the model studied, the simulation, as well as the measurements, are provided in Section II. Our findings related to the dynamical and structural properties of the thermal assembly of two-dimensional ring polymers are discussed in Section III. Finally, we provide a concluding discussion in Section IV.

\section{Modeling and Methods}\label{SEC-II}

\subsection{Model}

The ring polymers are modelled in the following manner. For the interaction between monomers, we use a combination of Weeks-Chandler-Andersen (WCA) potential
\begin{equation}
    U_{LJ}(r)=
    \begin{cases}
    4\epsilon[(\frac{\sigma_{m}}{r})^{12}-(\frac{\sigma_{m}}{r})^{6}]+\epsilon & \text{if }r\le 2^{\frac{1}{6}}\sigma_{m}\\
    0 & \text{if }r>2^{\frac{1}{6}}\sigma_{m}
    \end{cases}
\end{equation}
between all monomers, and finite extensible nonlinear elastic (FENE) potential
\begin{equation}
    U_{FENE}(r)=-\epsilon k_{F}R_{F}^{2}\ln [1-(\frac{r}{R_{F}\sigma_{m}})^{2}] \quad \text{if }r<R_{F}\sigma_{m}
\end{equation}
between bonded monomers, where $\sigma_{m}$ is the diameter of each monomer (and the unit of length), $\epsilon$ is the unit of energy, $k_{F}=15$ is the spring constant, and $R_{F}$ is the maximum extension of the bond ~\cite{smrek2020active}. 
For our study, we consider a system of  $N = 1000$ polymer rings, each consisting of $n_{m} = 20$ monomers. 
To avoid crystallization, we consider a 60:40 binary mixture by constructing two types of rings, with the {\em larger} ones having $\sigma_{m} = 1.0$, $R_{F}=1.5$
and the {\em smaller} ones having $\sigma_{m} = 0.7$, $R_F=1.0$. 
To model the flexibility of the ring, we have used an angular potential
\begin{equation}
    V(\theta)=K_{\theta}(1-cos(\theta-\pi))
\end{equation}
where $K_{\theta}$ is the angular stiffness. In our study, we consider rings of varying stiffness, viz. $K_{\theta}=10,20,50,100$.
In our study, we investigate how the variation of angular stiffness changes the dynamical and structural properties of the thermal assembly of ring polymers.

\subsection{Methods}

We initialized the system with a very low ring number density of 0.004. The initial placement of the particles is done randomly on a 2D square lattice, ensuring no overlap between the rings. To obtain glassy states, we work at a low temperature of $T=0.01$. We compress the system from a low-density (and therefore low-pressure) state to target higher pressure states by coupling it to the Nosé-Hoover barostat. After the system's energy stabilizes at the target pressure, we switch to the NVT ensemble to study the dynamics and structure at the average density corresponding to the targeted pressure; during the NVT runs, the pressure fluctuates around the value obtained via NVT.  We study the response of the assembly of ring polymers to two different rheological protocols. In one case, we impose an external shear rate ($\dot{\gamma}$), i.e., the simulation box is deformed continuously with a finite velocity. In the other case, we impose an oscillatory shear $\gamma(t)=\gamma_0\sin(2\pi{t/\tau_p})$, where $\gamma_0$ is the maximum strain amplitude during the cycle and $\tau_p$ is the time period of the oscillation. In this study, we fix $\tau_p=1000$ and explore the response to a range of imposed $\gamma_0$; with these choices, we explore shear rates in the range of $2\pi\gamma_0/\tau_p \approx 3\times10^{-4} - 3\times{10^{-3}}$ which is comparable to previous studies \cite{yeh2020glass}. During shear, the thermalization at $T=0.01$ is maintained via a Langevin thermostat (having dissipation timescale of $\tau_D=1$), coupled to the velocity component ($v_y$) transverse to the direction of applied shear ($x$).

The molecular dynamics (MD) simulations were conducted using LAMMPS \cite{plimpton1995fast}, wherein the inertial equations of motion were numerically integrated using a timestep of $0.005$. All quantities reported in this study are expressed in reduced LJ (Lennard-Jones) units. For averaging, we consider five independent trajectories, for each $K_{\theta}$ value.

\subsection{Observables}

%\subsubsection{Dynamics}

For characterizing the dynamical behaviour of the system, we measure the mean-squared displacement (MSD) of the centers of mass (c.o.m) of the rings. The MSD is defined as $\Delta r^{2}(t)=\frac{1}{N}\sum_{i=1}^{N}|\Vec{r_{i}}(t)-\Vec{r_{i}}(0)|^{2}$, where $N$ represents the total number of rings, and $\Vec{r_{i}}(0)$ and $\Vec{r_{i}}(t)$ denote the coordinates of the c.o.m. of the $i^{\text{th}}$ ring at times 0 and $t$, respectively. For systems in equilibrium, we calculate the ensemble-averaged MSD, $\langle\Delta r^{2}(t)\rangle$. From this, the diffusion coefficient $D$ is determined using $D=\lim_{t\to\infty}\frac{\langle\Delta r^{2}(t)\rangle}{4t}$, where $\langle{\cdots}\rangle$ indicates averaging over independent trajectories in our case.

%\subsubsection{Structure}
The shape changes of the rings can be quantified by measuring the asphericity and examining its evolution with imposed drive or thermal fluctuations. To calculate the asphericity of a ring, we use the gyration tensor, denoted as $S_{mn}$. The gyration tensor is computed using the formula: $S_{mn} = \frac{1}{N}\sum_{i=1}^{N}(r_{m}^{(i)}-r_{m}^{(CM)})(r_{n}^{(i)}-r_{n}^{(CM)})$, where $m$ and $n$ are Cartesian coordinate indices. In two dimensions, each ring corresponds to a gyration tensor with two eigenvalues, $\lambda_{1}^2$ and $\lambda_{2}^2$. The asphericity of a ring, denoted as $a$, can then be defined as follows:
%\begin{equation}
$a= [{(\lambda_{1}^{2}-\lambda_{2}^{2})^2}]/[{(\lambda_{1}^{2}+\lambda_{2}^{2})^2}]$.
%\end{equation}

\begin{figure}[h]
 \centering
    \includegraphics[width=0.9\columnwidth]{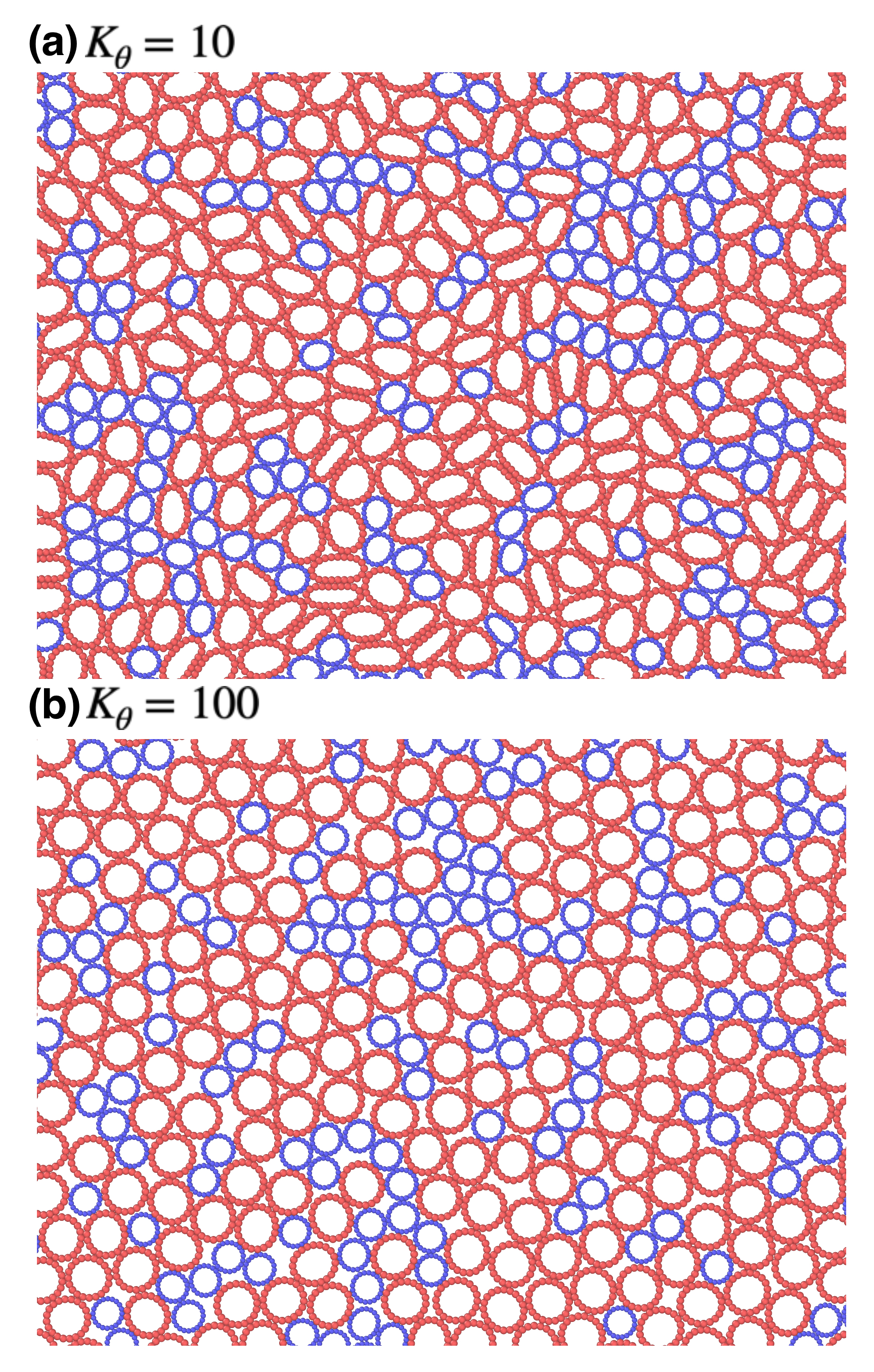}
    \caption{Zoomed in representative snapshots of the quiescent glassy states for different ring stiffness values at pressure $P=0.75$:(a)$K_{\theta}=10$ and (b)$K_{\theta}=100$. The larger and smaller rings are colored as red and blue respectively.}
    \label{snapshots}
\end{figure}

\begin{figure}
 \centering
 \includegraphics[width=0.9\columnwidth]{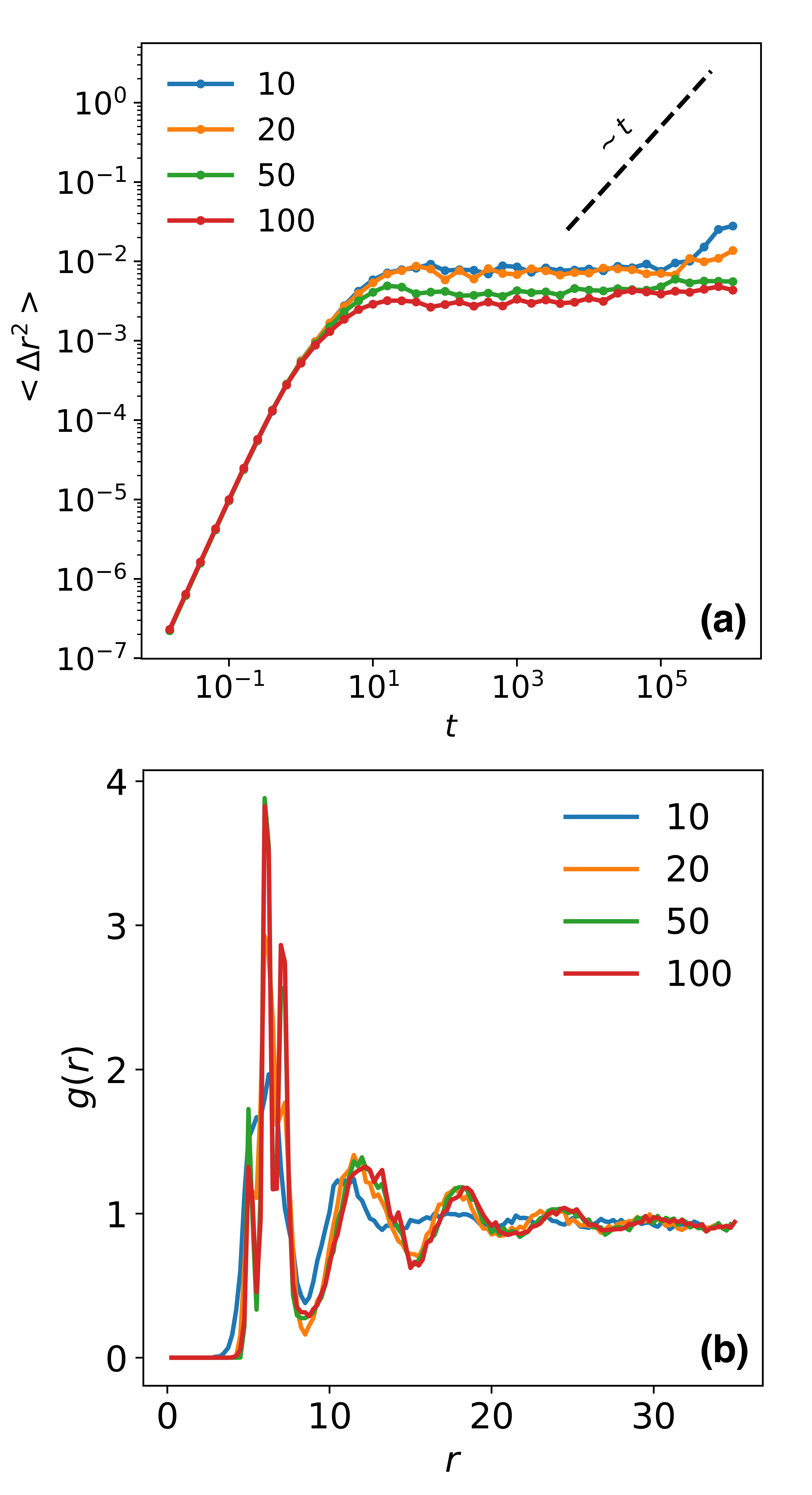}
 \caption{$P=0.75$. In the quiescent state, for the different stiffness values ($K_{\theta}=10, 20, 50, 100$) of the ring polymers, (a) time evolution of the mean squared displacement of the center-of-mass of each ring. and (b)  pair correlation function, $g(r)$, computed over all the rings. In (a), the dashed line corresponds to a linear dependence on time characteristic to diffusive dynamics.}
 \label{glass1}
\end{figure}

\section{Results}\label{SEC-III}
\subsection{Quiescent state}

The representative snapshots of the states obtained after compression to $P=0.75$, for the contrasting stiffness values that we have explored, viz. $K_{\theta}=10, 100$, are shown in Fig.\ref{snapshots}, and similar snapshots for $K_{\theta}=20, 50$ and for $P=0.75$, are shown in Fig S1 (see SI). For $P=0.50$, the corresponding plot is shown in Fig S2 (see SI). The non-circular shapes due to more flexibility are visible in the case of $K_{\theta}=10$, whereas the deviation from circular shape is negligible for $K_{\theta}=100$. These observations align with previous studies demonstrating how varying polymer stiffness can significantly influence the structural organization and dynamics of ring polymers in dense systems \cite{cai2022conformation,kruteva2023topology,roy2022effect,michieletto2017glassiness,gnan2019microscopic}. 
We also note that although these states are highly dense, they are still non-confluent; however, the assembly of more flexible rings is better at space filling than the more rigid ones due to the ability to change shape.

We assess the glassiness of these states by measuring the mean squared displacement (MSD) of the center of mass of the rings; see the top panel of Fig.\ref{glass1} for the data corresponding to the different stiffness values that we have explored, viz. $K_{\theta}=10, 20, 50, 100$ for $P=0.75$. This is also the case for $P=0.5$ as seen in Fig. S3 (see SI).  Within the time window of our observation, the MSD in all cases displays a plateau at long times.
This indicates dynamical arrest, which is a hallmark of glassy dynamics, where particles become trapped in cages formed by their neighbors, limiting their mobility — a phenomenon widely recognized in the study of disordered systems \cite{angell1995formation, weeks2000three}. The arrest of the dynamics due to caging, observed for all $K_{\theta}$ values, confirms that in each case the system is in a glassy state at our working temperature and pressure. This behavior is consistent with previous findings in dense polymer systems where increasing stiffness leads to a decrease in the vibrational motion of the constituent particles, as stiffer polymers are more constrained \cite{ness2017nonmonotonic}.
\begin{figure}
\centering
\includegraphics[width=0.9\columnwidth]{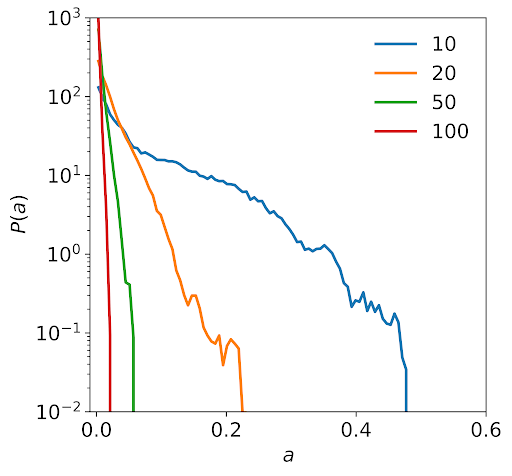}
\caption{$P=0.75$ Distribution of asphericity in the glassy state, for the different ring polymer assemblies having $K_{\theta}$ values as marked. The mean asphericity, for $K_{\theta}=10,20,50,100$, is respectively $0.075,0.020,0.005, and 0.0016$.
}
\label{glass2}
\end{figure}

To confirm the disordered nature of the obtained thermal assembly of the ring polymers, we compute the pair correlation function, $g(r)$, using the center of mass of the rings. The corresponding data, averaged over the rings as well as the ensemble of independent trajectories, are shown in the bottom panel of Fig.\ref{glass1}. The lack of long-range order in $g(r)$ indicates that the obtained structures are indeed disordered, a characteristic typical of glassy systems \cite{berthier2011dynamical}. Notably, with decreasing ring stiffness, the structures become more disordered, as reflected by the more pronounced differences in the shape of $g(r)$ between the pair correlation functions for $K_{\theta}=100$ and $K_{\theta}=10$.  We further characterize the quiescent glassy state by analyzing the shapes of the rings under ambient pressure for the different stiffness values. The ring shape is quantified via the asphericity. In Fig.\ref{glass2}, we show the distribution of asphericity computed in the glassy state for different values of $K_{\theta}$. As expected, for $K_{\theta}=100$, the distribution is very narrow, indicating that the rings maintain a nearly circular shape with little variation, consistent with the behavior of stiff polymers. As the stiffness decreases, the distribution of asphericity broadens, with the assembly of rings having $K_{\theta}=10$ exhibiting a fat-tailed distribution that reflects a greater diversity in ring shapes. This broad distribution suggests that softer rings are more susceptible to shape fluctuations and explore larger conformational fluctuations. The contrasting variations in shape across the changing stiffness of the rings are clearly visible in the representative snapshots shown in Fig.\ref{snapshots}.

\subsection{Steady state rheology: Imposed shear-rate}
\begin{figure}
 \centering
    \includegraphics[width=0.9\columnwidth]{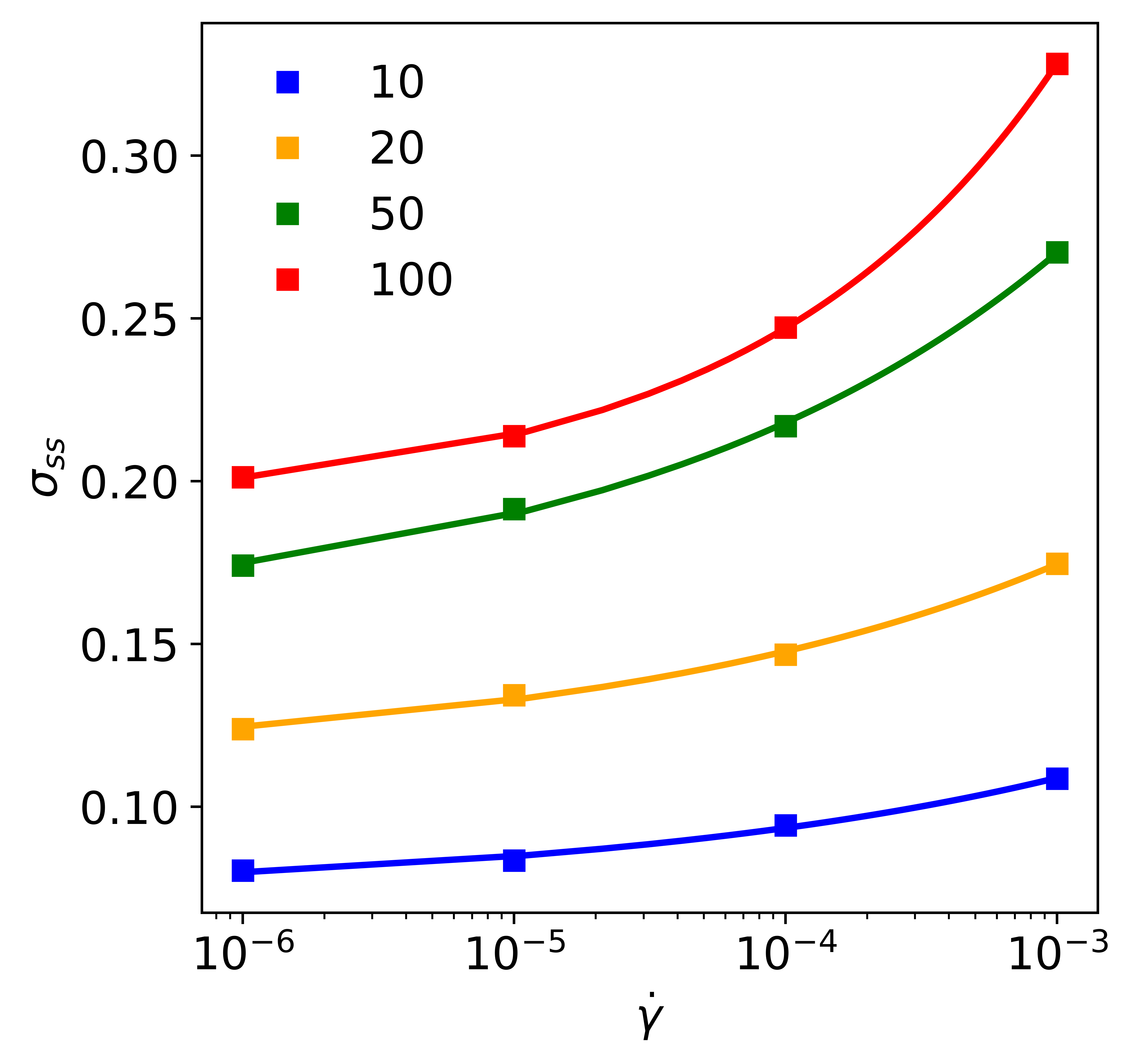}
    \caption{$P=0.75$. For different $K_{\theta}$ values as marked (shown in squares), the dependence of steady state shear stress ($\sigma_{ss}$) on imposed shear-rate ($\dot{\gamma}$). In each case, the lines correspond to Herschel-Bulkley fits to the data, to obtain the yield stress (see text for details).}
 \label{ssflow}
\end{figure}

To further establish the glassiness of the states obtained at $P=0.75$, we rheologically probe the rigidity by measuring the steady-state response to an externally imposed shear rate ($\dot{\gamma}$). We measure the steady-state stress for a range of imposed $\dot{\gamma}$; the data is shown in Fig.\ref{ssflow} for different values of ring stiffness. The data are well-fitted by the Herschel-Bulkley function $\sigma=\sigma_y+A\dot{\gamma}^\beta$, where $\sigma_y$ is the yield stress and $\beta$ is the Herschel-Bulkley exponent~\cite{bonn2017yield}. The obtained yield stress values are $\sigma_y=0.192, 0.157, 0.124, 0.073$ for $K_{\theta}=100, 50, 20, 10$, respectively, with $\beta$ in the range $0.25-0.39$. The existence of finite yield stress in each case demonstrates that the bidisperse mixture of ring polymers, for all stiffness values, behaves as a yield stress material, which is a characteristic feature of glassy systems \cite{fielding2000aging, varnik2003shear}. As expected, with decreasing stiffness, the yield stress of the material decreases, indicating increased softness and more pronounced structural rearrangements under shear \cite{berthier2011theoretical}. This trend is consistent with previous observations in other glass-forming systems, where softer materials typically exhibit lower yield stresses due to their enhanced ability to undergo deformation at lower applied stresses \cite{liu1998jamming, bi2011jamming, meijer2005mechanical}. The variation in yield stress with stiffness, as observed in our study, highlights the influence of intrinsic polymer properties on the development of squishy mechanical response of glassy materials having deformable constituents.

\subsection{Response to oscillatory shear}
\begin{figure}
 \centering
    \includegraphics[width=0.8\columnwidth]{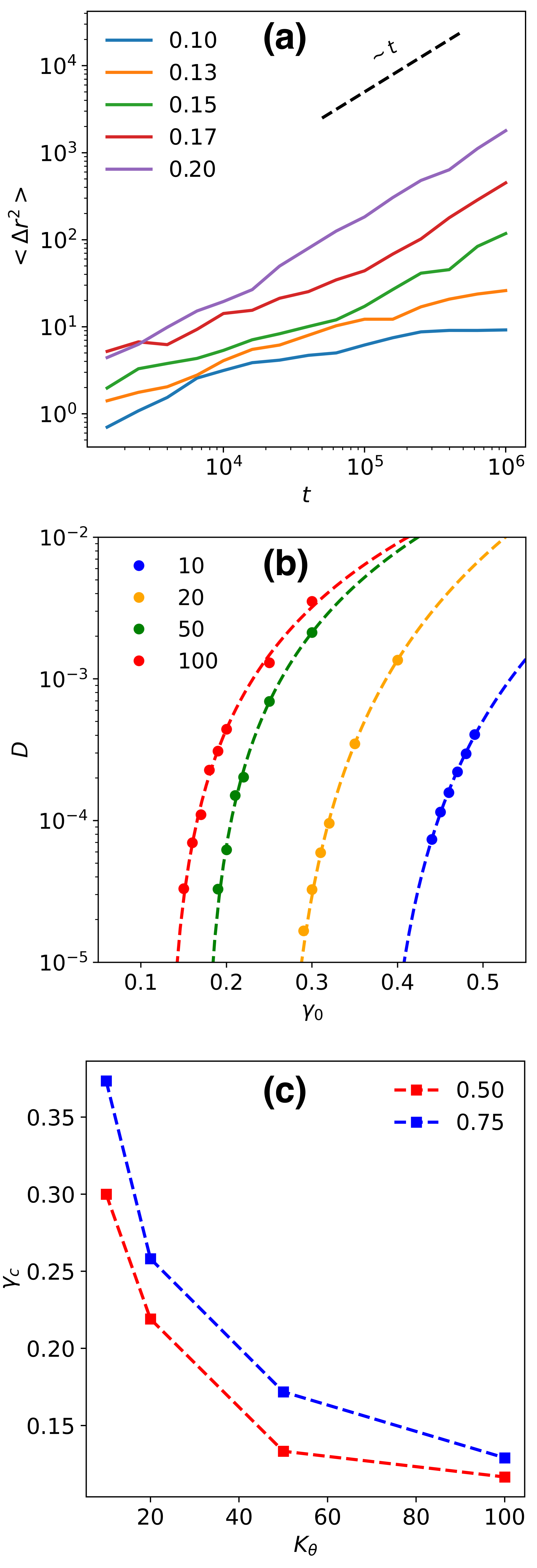}
    \caption{(a) $P=0.75, K_{\theta}=100$. Time evolution of center of mass MSD, for different $\gamma_0$, measured stroboscopically, relative to the unsheared glassy state. (b) $P=0.75$. Variation of diffusion coefficient extracted from MSD with $\gamma_0$, for different $K$. Data fitted by power law to extract yield threshold $\gamma_{c}$, for different $K$. (c) Dependence of  $\gamma_c$ on $K_{\theta}$ for two different $P$ values.}
 \label{yieldflow}
\end{figure}

It is well-known that a soft amorphous solid loses rigidity and fluidizes when the strain amplitude of the imposed oscillatory shear exceeds a threshold value \cite{bonn2017yield, coussot2007rheophysics}. Hence, such a yielding response is also expected for the glassy states of ring polymers.  To identify the threshold yield strain, we monitor the mean squared displacement (MSD) of the center of mass of the ring polymers across stroboscopic frames (i.e., when the global shear strain is zero during the oscillatory cycle), relative to the initial undeformed state. In the fluidized state, the dynamics is typically diffusive, with the diffusion coefficient increasing as $\gamma_0$ increases \cite{koumakis2013complex, edera2024yielding}. Below the yield threshold, the dynamics should remain glassy, meaning the rings either stay localized or exhibit sub-diffusive behavior.  We show the example of  $K_{\theta}=100$ in the top panel of Fig.\ref{yieldflow}(a). We note that for small $\gamma_0$, the caging lengthscale increases with increasing strain amplitude, indicating the increased vibrations with increasing shear. From the MSD data at large $\gamma_0$, we extract a diffusion coefficient, $D$, and present the corresponding data for different $K_{\theta}$ values in Fig.\ref{yieldflow}(b). The corresponding figure for $P=0.50$ is shown in Fig. S4 (see SI). The dependence of the diffusion coefficient on the strain amplitude $\gamma_0$ is well-fitted by a power-law $(\gamma_0-\gamma_c)^n$, where $\gamma_c$ is the estimated threshold strain amplitude \cite{fiocco2013oscillatory}. This threshold, $\gamma_c$, represents the strain magnitude below which the amorphous solid ceases to fluidize, i.e. retains its rigidity. We observe that this threshold depends on the ring stiffness. For the rigid rings, $\gamma_c \approx 0.13$, which is consistent with previously reported thresholds for atomistic amorphous solids or hard sphere colloidal glasses \cite{yeh2020glass, koumakis2013complex}. With decreasing $K_{\theta}$, $\gamma_c$ increases; for $K_\theta=10$, the estimated threshold is approximately $0.37$, which implies that the strain range for the rigidity is enhanced by a factor of around $3$. The increased flexibility in the rings thus allows them to better accommodate the applied shear \cite{poincloux2024squish}, thereby maintaining rigidity until the developing stress is no longer sustainable by the disordered structure, at which point the solid yields. We studied the response at two different ambient pressures, $P=0.5$ and $P=0.75$, with the locus of $\gamma_c(K_{\theta})$ for both cases shown in the bottom panel of Fig.\ref{yieldflow}. At lower pressure, the solid is less rigid, and for any given $K_{\theta}$ value, the strain amplitude $\gamma_c$ is correspondingly smaller. Notably, for the more rigid rings ($K_{\theta}=100$), $\gamma_c$ shows little variation with pressure, suggesting a robustness of the yield threshold in these systems under compaction. 

\begin{figure}
 \centering
    \includegraphics[width=\columnwidth]{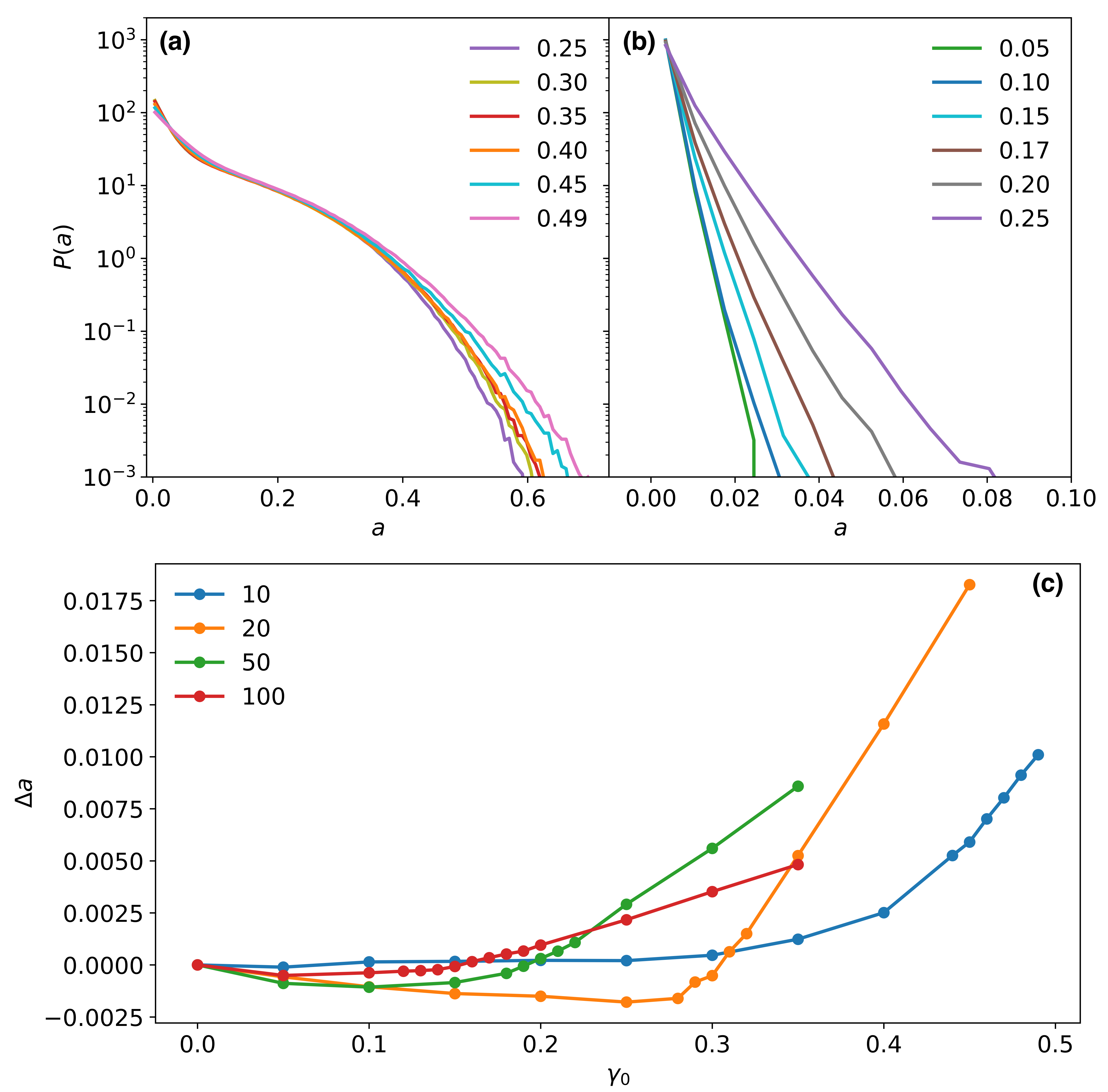}
    \caption{Steady State distribution of asphericity of the rings, as a function of strain amplitude $\gamma_0$ for $K=10$ (a), $100$ (b). (c) Change in mean asphericity of the rings $\Delta{a}$, relative to the unsheared state,  with varying strain amplitude, for the different stiffness values as marked.}
 \label{shapes}
\end{figure}

Next, we examine the interesting question of how the shape of the rings is affected by the applied shear. In the top panel of Fig.\ref{shapes}, we present the steady-state distribution of asphericity as a function of $\gamma_0$ for the contrasting cases of $K_{\theta}=10$ (a) and $100$ (b). In Fig.\ref{shapes}(c), we show how relative to the unsheared state, the mean asphericity changes ($\Delta{a}$). A corresponding figure for $P=0.50$ can be seen in Fig.~S5 (see SI). Prior to yielding, the mean shape remains nearly constant or anneals, relative to the unsheared glassy state. However, beyond the yield point, the distribution shifts, and the mean asphericity increases steadily with $\gamma_0$. Further, from the distributions, it is observed that primarily the change in shape occurs for rings that are already non-circular, i.e. the tail of the distribution fattens with increasing $\gamma_0$. The observed increase in asphericity with shear is consistent with the expected behavior of polymers under deformation \cite{hebraud1998mode}, where the rings stretch and deviate more from their circular shape as the applied strain increases .

\begin{figure}
	\centering
	\includegraphics[width=0.9\columnwidth]{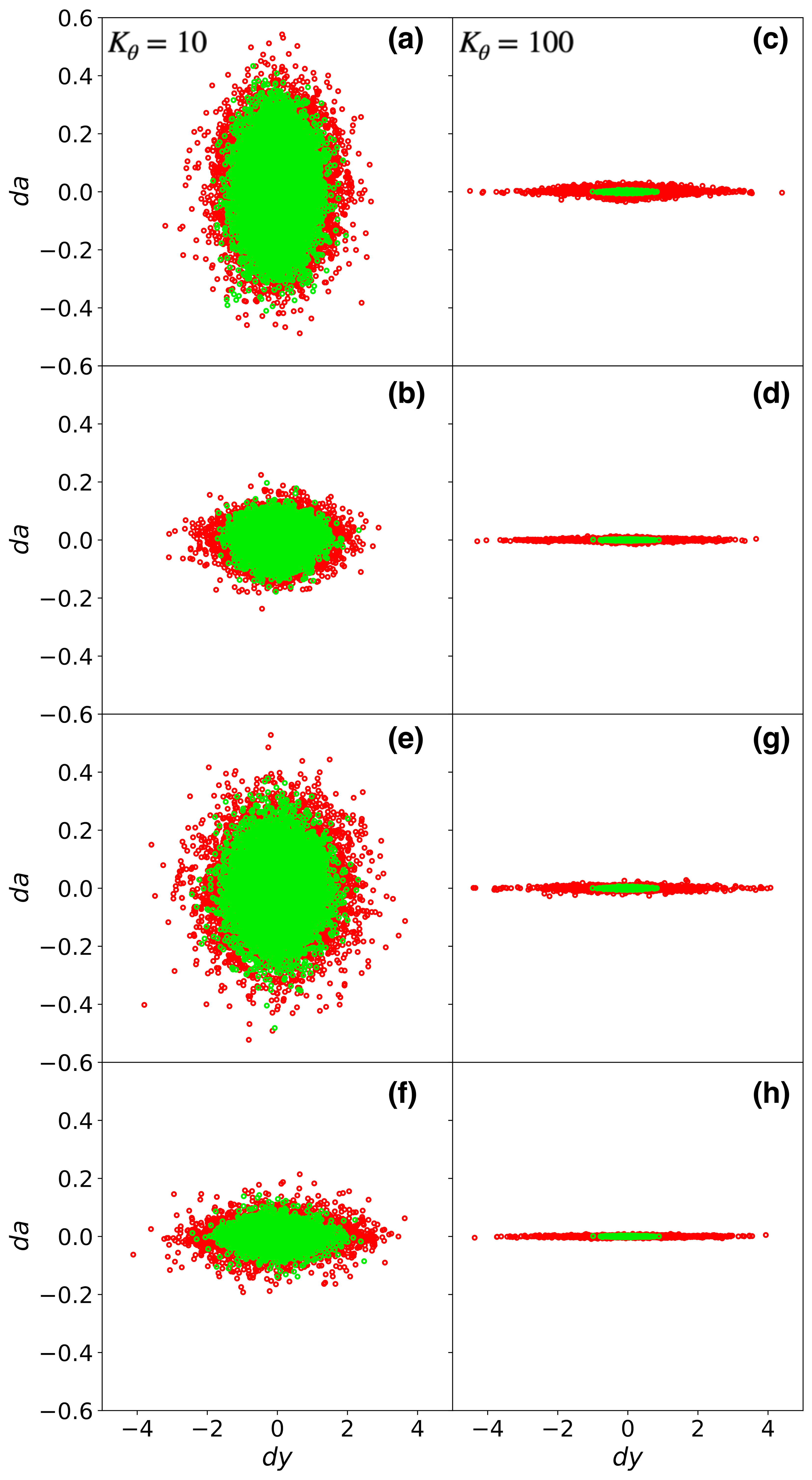}
	\caption{
		$P=0.75.$ Interplay of shape change and non-affine motion during shear cycle, for the most stiff ($K_\theta=100$) and most flexible ($K_\theta=10$) rings, computed at global strain of (a)-(d) $\gamma_0$ (quarter-cycle) and (e)-(f) zero (half-cycle). Differences are calculated relative to the global zero-strain frame preceding the quarter or half cycle. For $K_\theta=10$, green scatter corresponds to $\gamma_0=0.30$ and red scatter corresponds to $\gamma_0=0.45$. For $K_\theta=100$, green and red correspond to $\gamma_0=0.05, 0.15$, respectively. Data for large rings are shown in (a),(c),(e),(g), and small rings are shown in (b),(d),(f),(h).}
	\label{cloud}
\end{figure}

Having established that shape changes are becoming more significant in the plastic flow regime, we next probe the interplay between the shape change of each ring and the non-affine motion of its center-of-mass within a shear cycle, to determine its influence in the ensuing plasticity. While the shape change is quantified via the change in asphericity ($da$), the non-affine motion is quantified via the displacement ($dy$) in the direction transverse to the applied shear ($x$). For this analysis, we consider two states: (i) where the applied global strain is maximum, i.e., at the end of each quarter cycle in the $\pm{x}$ direction, and (ii) where the applied global strain reaches zero during the cycle, i.e., at the end of each half cycle. In each case, the deviations in shape or position are measured relative to the zero-strain frame occurring immediately before (i) or (ii). We probe this interplay between shape and motion for the two thermal assemblies of contrasting stiffness, viz. $K_{\theta}=10, 100$. The data gathered over several cycles, in the steady state, is plotted in Fig.\ref{cloud}, where we compare the response of the large and small rings in the elastic (shown in green scatter) and plastic (shown in red scatter) regimes.

In a perfectly elastic process for rigid objects, the particles would undergo affine translation during a quarter cycle and recover their original location at the end of the half cycle. For deformable objects, a similar behavior would occur for the shape degree of freedom, where the object would get stretched during the quarter cycle and recover shape at the end of the half cycle. However, in the presence of thermal noise, the reversibility over the half-cycle is, of course, not perfect, even in the elastic regime \cite{hima2014experimental}. Both these aspects are visible in the green scatter in Fig.\ref{cloud}; for both $K_\theta=10$ and $100$, we have chosen a strain value for which glassy dynamics is observed, viz. $\gamma_0=0.3, 0.05$ respectively. For the flexible rings, there is extensive shape change, especially for the larger rings, during the quarter cycle (see (a), (b)), as well as non-affine motion ($dy$). Some of the translational and shape changes are recovered at the end of the half cycle (see (e), (f)). For the stiffer rings, the scale of shape or translational change is comparatively much less; see (c), (d), (g), (h). 

If we now focus on the plastic regime (shown in red scatter), for ${K_{\theta}=100, \gamma_0=0.15}$, both large and small rigid rings undergo non-affine motion of similar magnitudes at the end of the half-cycle; see (g), (h). In contrast, for ${K_{\theta}=10, \gamma_0=0.45}$, it is the smaller rings that undergo more non-affine motion, but the magnitude of that is less than compared to the $K_{\theta}=100$ rings. On the other hand, the larger $K_{\theta}=10$ rings exhibit more irrecoverable shape changes. Also, a key observation is that rings undergoing large shape change ($da$) within half-cycle undergo least displacement ($dy$), whereas those undergoing large non-affine motion undergo least shape change, which is consistent with the recent experimental findings \cite{poincloux2024squish}. Thus, these observations shed light on the nature of plasticity occurring within the $K_{\theta}=10$ assembly, where there is an interplay between the shape change of larger rings and the non-affine motion of smaller rings, leading to structural changes and flow.

\begin{figure}
 \centering
    \includegraphics[width=0.9\columnwidth]{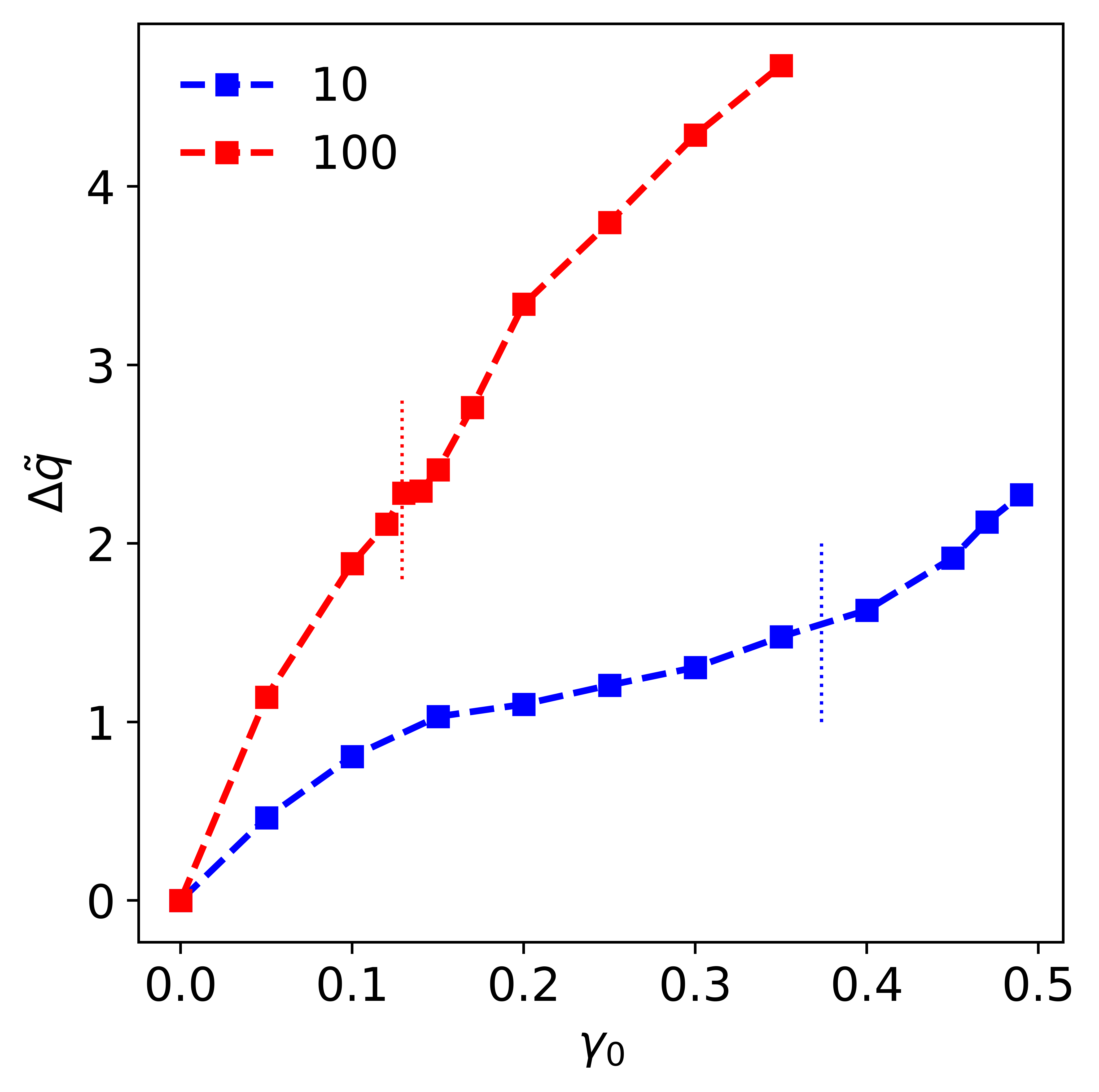}
\caption{$P=0.75.$ For rings of different stiffness ($K_{\theta}=10, 100$), average change in the number of contacts of a ring with neighbouring rings, at the monomer scale, with varying strain amplitude $\gamma_0$; $\Delta{\tilde{q}}$ is the measure relative to $\tilde{q}(\gamma_0=0)$. The ensemble averaged yield threshold for $K_{\theta}=10, 100$ are also marked by dashed vertical lines.}
\label{qtilde}
\end{figure}

Finally, we demonstrate that even though the network formed by the center-of-mass of the rings remains largely intact for $\gamma_0 < \gamma_c$, there are changes in local contacts between neighboring rings. For this analysis, we focus on the contact between monomers of one ring and the monomers of neighboring rings and examine whether there are changes in the average number of such contacts between consecutive stroboscopic frames. We take each ring $i$ and mark all the neighboring rings $j$. If, between two consecutive frames, the change in the number of monomers of $i$ in contact with monomers of $j$ is $\Delta n_{ij}$, then the average extent of change in contact of each ring between two consecutive frames is quantified by $\tilde{q}= [\sum_{i} \sum_{j} |{\Delta n_{ij}}|]/N$. In a dense thermal packing of rings, changes in such local monomer-monomer contacts can either happen if there is a relative slip between neighboring rings, even when the overall ring network does not change, or when there is a plastic event such that the two rings are no longer neighbors. Since there are few such events in the quiescent state due to thermal fluctuations, we quantify the change brought in by shear via $\Delta\tilde{q}=\tilde{q}(\gamma_0)-\tilde{q}(\gamma_0=0)$. In Fig.\ref{qtilde}, for the contrasting cases of $K_{\theta}=10, 100$, we plot $\Delta\tilde{q}$ as a function of $\gamma_0$, and also mark the respective locations of $\gamma_c$ with vertical lines. For $K_{\theta}=100$, there is a finite $\Delta\tilde{q}$ below $\gamma_c$ which mainly comes from rolling of one ring over another, and this number grows rapidly with $\gamma_0 > \gamma_c$, when plastic events proliferate. For $K_{\theta}=10$, however, the growth is weaker before $\gamma_c$ and then rapidly increases when large-scale plasticity sets in. The slow growth in the case of more flexible rings, when yielding has not occurred, suggests the occurrence of intermittent slippages due to the applied shear. This gives us a sense of the effective friction at play between the soft, deformable objects driven by the external shear, which needs further investigation.

\section{Discussion and Conclusion}\label{SEC-IV}

Understanding yielding and flow in soft squishy matter has significant practical implications for various materials, including foams, gels, emulsions, and biological tissues, where controlling the transition between fluid-like and solid-like states is crucial for material design and functionality~\cite{weeks2007soft,vlassopoulos2014tunable,reichhardt2014aspects,behringer2018physics,cohen2014rheology,ji2020interfacial,van2009jamming,pan2023review}. In this study, we focused on the yielding transition in a highly dense, bidisperse 2D system of deformable polymer rings under oscillatory shear and examined how varying the stiffness of these rings influences their dynamical and structural properties. We modeled the system using a combination of the Weeks-Chandler-Andersen (WCA) potential for non-bonded interactions and a finite extensible nonlinear elastic (FENE) potential for bonded monomers, applying an angular potential to account for the flexibility of the rings. We introduced bidispersity through different WCA coefficients and FENE bond lengths to prevent crystallization. 

We first demonstrated that at the selected phase space of ring polymer stiffness and pressure values, the thermal assembly of rings exhibited glassy dynamics, at the state points of our choice, by monitoring the mean squared displacement of the center-of-mass of the rings, after compression to the targeted density. The pair correlation function further confirmed the disordered nature of the system, with disorder becoming more pronounced at lower stiffness values. The distribution of asphericity, which measures ring shape, varied with ring stiffness, showing a broader distribution for softer rings, indicating greater shape fluctuations. We then investigated the steady-state rheological response of the system to an imposed shear rate. Our results demonstrated yield stress behavior across all stiffness values, a characteristic of glassy systems. The yield stress decreased with decreasing stiffness, thus becoming more soft and squishy for the smallest ring stiffness that we have studied.

In response to oscillatory shear, the system exhibited a fluidization transition when the strain amplitude exceeded a threshold value, with this threshold depending on ring stiffness. The assembly of softer rings (lower stiffness) maintained rigidity for a large window of strain amplitudes (nearly three times the most rigid ones) but eventually yielded. The distribution of asphericity under shear showed that the rings became increasingly deformed as the strain amplitude increased, particularly beyond the yield point; the rings that were more aspherical in quiescent state underwent more shape change under shear. We observed that the increase in the yield threshold with decreasing stiffness suggests a nuanced interplay between particle deformability and shear-induced fluidization, with the smaller rings within the mixture exhibiting larger non-affine motion and less shape change while the larger rings display larger irreversible shape deformations but less translational plasticity.  Further, we also analysed local contact changes, which occur even below the yielding limit, suggesting occurrence of rolling and sliding which thus leads to effective friction between soft deformable particles arising out of the spatial details encoded into the ring modeling.  These findings underscore the complex interplay between particle stiffness, deformation, and mechanical response in disordered systems, offering new insights into the jamming transition and the behavior of soft colloids under varying conditions.

Future studies will explore in greater details the local plasticity in the softer ring polymeric system that builds up to large-scale flow. Further, the role of the shearing frequency as well as microscopic dissipation needs to be investigated to understand the complexities at play. Eventually, by tuning the ring properties as well as the mixture composition, the aim would be to design bio-mimicked systems that can lead up to functionalized materials using such soft deformable objects.

\section*{Acknowledgment}
 We thank the HPC facility at the Institute of Mathematical Sciences for computing time. PC and SV also acknowledge support via the sub-project on the Modeling of Soft Materials within the IMSc Apex project, funded by Dept. of Atomic Energy, Government of India.

%\section*{Data Availability}
%The data that support the findings of this study are available from the corresponding author upon reasonable request. 

\bibliography{2DSquishy}

\end{document}